# The Virtual Machinery Workshop: an Immersive 360º Web-Based Vocational Education Learning

Prayitno, Anwar Sukito Ardjo, Liliek Triyono, Bambang Kuswanto.
Department of Electrical Engineering, Politeknik Negeri Semarang, Indonesia
prayitno@polines.ac.id

**Abstract** In this article, we have proposed an educational learning material model using 360-degree environment on web-based platform by personalizing the learning environment as per user surf the web. Virtual Machinery Workshop allow an illusionary immersion into a virtual learning environment where the user can freely navigate through a 360-degree videos and pictures. The main objective of this study is to analyze the nature of the added information in virtual learning environment, cognitive difficulty of user to study at web-based learning and the potential immersive learning in education. Our proposed method presented machinery workshop situation by utilizing 360-degree pictures and videos technique. The results show that our learning environment enhancing users study information through this visualized system.

## I. INTRODUCTION

Facing globalization of education and industrial revolution 4.0, the government through the ministry of technology research and higher education must be aware of disruption technology era. The era of disruption technology has brought challenges for education, especially the ability of students as prospective graduate needs to be improved with digital and innovative technology. To face this era, learning strategy as e-learning content and Augmented Reality form are needed. These technologies are in line with 21[st] century learning method with student as centered learning.

21[st] Century learning method enhance the student to access, manage, create and share information through many media technologies. The commonly teaching and learning method in classroom, sometimes students start to be not interested and occasionally do not know the learning topic they learn. Some researchers indicate that high quality teaching and learning require active learning activity from the students. Active learning can occur when students participate in discussion by 70%, while practicum either in the workshop or in simulation can increase to 90%.

The process of meaningful teaching and learning has several challenges for vocational education university, especially process of learning experience in polytechnic as vocational colleges emphasizes the ratio between practicum hours is greater than the hours of learning theory. Students are emphasized to feel more hours in the workshop to carry out lab work. Optimal supplementary material in machinery curriculum is more challenging than other subject, because outside the lab class hour, student need to be enhanced with lab environment such as machinery tools, equipment and tutorial. Recently, several research [1]–[4] utilizing immersive technology to enhance education quality. Immersive technology allow user see real world scene from visuals and multimedia information through computer machine. Virtual Reality (VR) learning experiences are engaging and allow students to immerse themselves in content beyond what is possible in the real-world.

In this paper, we aim at applying a variation of immersion technology to create a meaningful teaching and learning process. But only immersive technology through 360-degree virtual environment meet our need to build for machinery workshop learning material. Firstly, by collecting workshop environment in 360-degree we build virtual learning environment. Then, we insert audio visual machinery learning material into virtual environment.

Our work is exhibited as follows:
- We propose a virtual machinery workshop based on 360-degree environment immersive technology
- We carry out web-based simulator to experiment our method and to verify the performance.
- We test the pedagogical effectiveness of 360-degree virtual learning experience.

The remainder of this paper is organized as follows. We introduce some related works in section II. Following in section III, we describe the details of our proposed learning application framework. In Section IV, we carry experiments and launch an analysis of the result. Finally, we conclude

This work was supported by the Politeknik Negeri Semarang Research Fund.

Anwar Sukito Ardjo is senior researcher at Politeknik Negeri Semarang. He is now with the Department of Machine. (e-mail: anwars.ardjo@gmail.com)
.

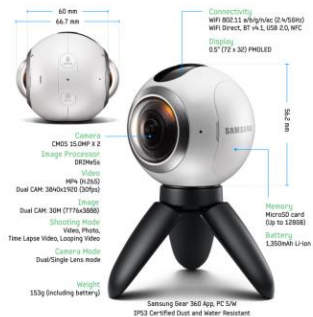 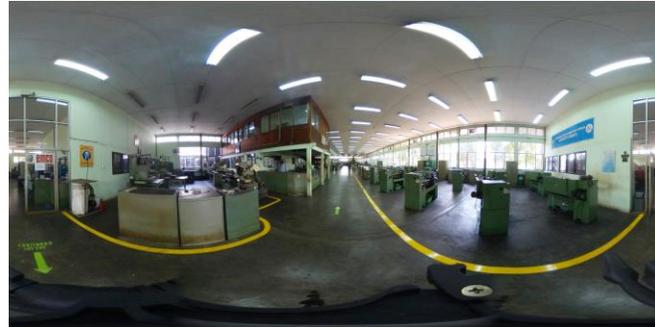

Figure 1. Left side represent 360-degree picture and video record tool and right side depict equirectangular projection result from camera

the paper and discuss our future work at section V.

## II. RELATED WORK

With rapid development hardware and software technology of web, virtual reality and smartphone in modern society, it can be utilized as entertainment as well as significantly for learning virtual environments through eLearning. Virtual environment is a way user interact with computer like in real-world intuitively. This virtual environment can be built by taking the real-world scene into pictures and videos in 360-degree. To make a tiled 360-degree virtual reality picture or video, one can resort to either multiple camera setup or partitioning of a single video into multiple frames of smaller resolution. An example of 360-degree picture and tools shown in Figure 1.

Recently, utilizing immersive media has gained increasing popularity among researchers in diverse area. There are several studies related to virtual reality conducted in health curriculum. For Example, Ouellet et al. [3] proposed implementation of virtual reality technology for the assessment of everyday memory, this research enhance neuropsychology learning health care material. Virtual shop helps young and adult learner build memory validity in daily routines. Similarly, Kleinert et al. [4] also presented a framework named an Immersive Virtual Patient Simulator (IPS) for training clinical decision in surgery. IPS prototype build with Artificial Learning Interface for Clinical Education (ALICE) simulator shows high students' acceptance for supporting daily clinical curriculum. In [6], Harrington et al. described the successful development and evaluation of a 360-degree operative video surgery operation to medically undergraduate students. This new video format demonstrated significant engagement and attractiveness benefit to traditional 2-dimensions format.

The concept of immersive media using virtual reality / 360º video is not a new approach, there are several studies also using this approach. Such as, Gundlach et al. [17] developed a tool for teacher to teach earth science curriculum. This immersive museum learning program simulator successfully attract students to learn earth science material. On the other hand, Zhihan Lv [18] introduced a framework to learn geography curriculum using digital glasses technology as interactive scene semantics. This technology provides 3-dimension virtual reality with global positioning system (GPS). The result of this research shows a promise future research in geographical big data.

There are several factors relevant to the effectiveness of Virtual Reality learning experiences. One is the level of immersion allowed by the technology. A highly immersive VR experience is one that provide sensory immersion. This means the outside world is shut out, reducing distractions away from the content, owing to the use of a head mounted display and headphones. Such an immersive virtual environment may lead learners to spend more time engaged with the content and decreases instances of mind-wandering. Another research shows that solid graphics and higher interactivity are key to effectively engaging learners with the content material. The 360-degree videos are omnidirectional panoramic videos that allow the viewers to pan and tilt in an uninterrupted circle rather than the fixed viewpoint of a traditional video. These videos can be viewed on phones by panning and tilting the phone, or on other devices, like low cost phone-enabled VR headset (e.g. Google Cardboard), or dedicated VR (e.g. Oculus Rift), by turning one's head similar to exploring the real world. Because viewers 360-degree videos have the agency to look around and explore different parts of the scene, these videos are more immersive than traditional 2-dimension videos, but less than a truly immersive VR learning experience.

Nevertheless, the use of VR technology in education is not without its challenges. For instances, the difficulty and time consuming of using VR applications as reported by Rentzos et al [15], particularly without well design interfaces.

## III. PROPOSED METHOD

The method to developed virtual machinery workshop application based on waterfall software development that have five phases such as problem analysis, design, implementation, testing and maintenance. A waterfall software development is a sequential software development process, where progress is seen as continuously flowing down (like a waterfall). Detail of each phases as follows:

*1) Problem Analysis*

This phase is more like to requirements engineering. Virtual machinery workshop simulator using 360-degree environment visual and multimedia is supplementary learning material for machinery curriculum. We proposed the system from mainly three aspects as shown at Figure 2 namely learning resource collection, virtual environment and learning material processing and web-based simulator application.

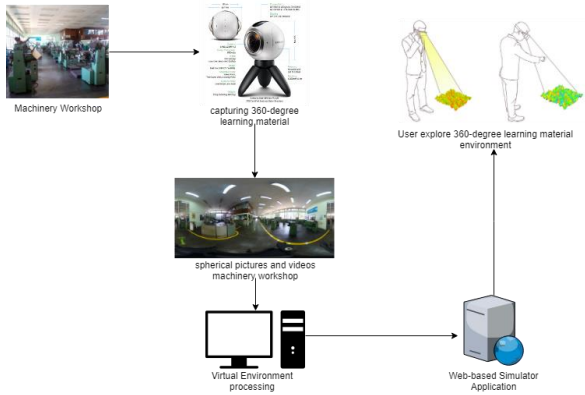

Figure 2. System Architecture

*2) Design*

Collecting learning material data was done at this design phase. There are three area at machinery workshop had been take for building virtual environment such as introduction area for beginner, medium area and advance area. First part is on the left side of machinery workshop scene. Left part scene is how the user will firstly enter the virtual workshop. The learning resource that could be collected in this part is the basic tools needed by the students. Second part of scene is middle of workshop scenes, at this scene users can explore about intermediate machine and equipment learning material. Finally, last scene part is the scene advance machine and equipment. We collected pictures and videos when teaching and learning activities conducted for next processing phase. The next material collected was manual description about tools and machines equipment was collected in form of text, pictures and videos.

Machinery workshop environment was recorded using a 360-degree dual lens camera with 4096 x 2048 (24fps) video resolution and 2304 x 1296 picture resolution. Pictures and video results collected by 360-degree camera are different than traditional camera. The output file is picture in equirectangular projection. Equirectangular projection represents the sphere as a 2x1 rectangle, where the cartesian X-Y coordinates correspond to the longitude and latitude of the point on the sphere. As a result, there's a lot of warping around the poles.

*3) Implementation*

In order to implement the virtual machinery workshop, first process was creating the virtual environment scenario and second process was inserting learning material as hotspot point. The final virtual workshop shown at figure 3.c where user can explore surrounding by clicking the picture and video detail icon. To move between object and scene, user tilting, panning or zooming virtual application.

A virtual environment learning scenario built by connecting each panorama pictures as shown at Figure 3.a. This learning path is like storyboard for the user. Scenario is a graphic organizer in the form of panoramic images displayed in sequence for the purpose of interactive media sequence. Furthermore, the user explores the environment written in this scenario. Before a picture result produce by 360-degree camera it must be processed into panorama

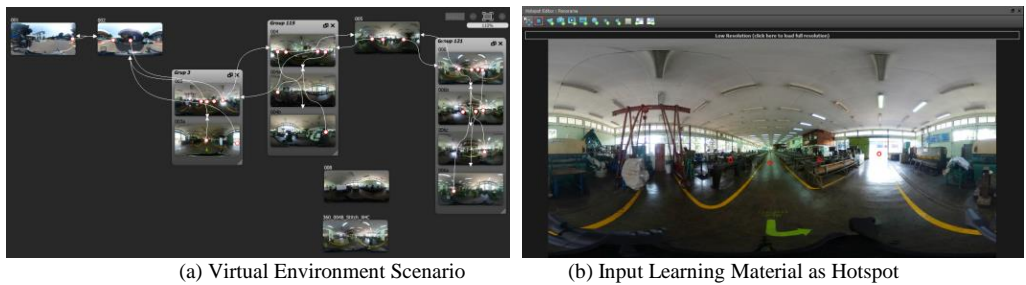

(a) Virtual Environment Scenario    (b) Input Learning Material as Hotspot

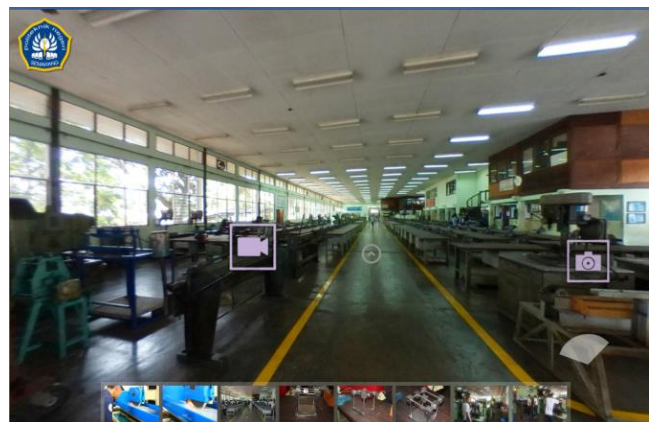

(c) scenario and hotspot integration
Figure 3. Virtual Machinery Workshop Learning material stitching.

picture or full spherical 360 degree. There are three requirements should be fulfilled to process properly such as the photo must have 2:1 aspect ratio, the Exif XMP tag value "ProjectionType=equirectangular" and they are below the maximum resolution and file size.

After completing the scenario to build virtual environment, then learning material such as texts, pictures and videos inserted to application. These learning materials inserted to application by hotspot point, a red circle in panorama image scene. This point will be shown as picture icon or image icon and located at specific tools or equipment to show the detail of its process.

*4) Testing*

The last part of the system shows virtual learning material into web based or virtual reality glasses to the user. Simulator application was uploaded to shared web server with apache 2.4 and PHP 5.6 specification. The performance testing was conducted using google chrome developer tools. The performance shown at Figure 4 depict load time around 0 – 6.51 seconds. It shows that the website simulator performance well. The scripting loading time show the most needed time to load into the web browser.

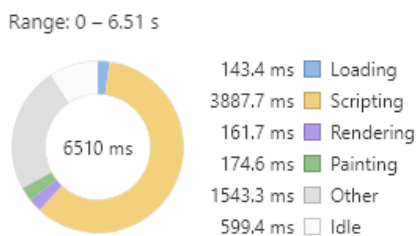

Figure 5. Website Load Performance

*5) Maintenance*

Maintenance of the system is the last part of waterfall software development method. In this part, revision and update has been done regarded to feedback from the user when they accessed the web-based application.

IV. RESULT AND DISCUSSION

Virtual machinery workshop application with immersive technology 360-degree view can be accessed through website at http://bengkel360.prayitno.web.id/ or through virtual reality tools as shown at figure 5. User can explore virtual workshop by tilting and panning surrounding scenes. At upper left figure 5 depict a virtual small world, it gave the user presence of real-world scenery. Then, at upper right figure 5 shown to the user first appearance of the application. This scene was at the front information of machinery workshop polytechnic state of Semarang, when usually people enter the building. Figure 5 lower left is the scene when user click the video icon. It shown a detail of machine in video format. Video was stream from the YouTube to save the space and maintain the bandwidth of web server. Finally, figure 5 lower right shown when user click the picture icon. This used for machine or tools detail in picture format.

Although the volume of 360° videos is likely to increase, further study into the potential information retention of the platform is required. As both engaging and immersive, this novel video platform delivers significant benefits to audiences and may appeal to modern learning styles. The 360° videos may add a new dimension to education including orientations to new environments, team training, and formal-examinations across multiple disciplines.

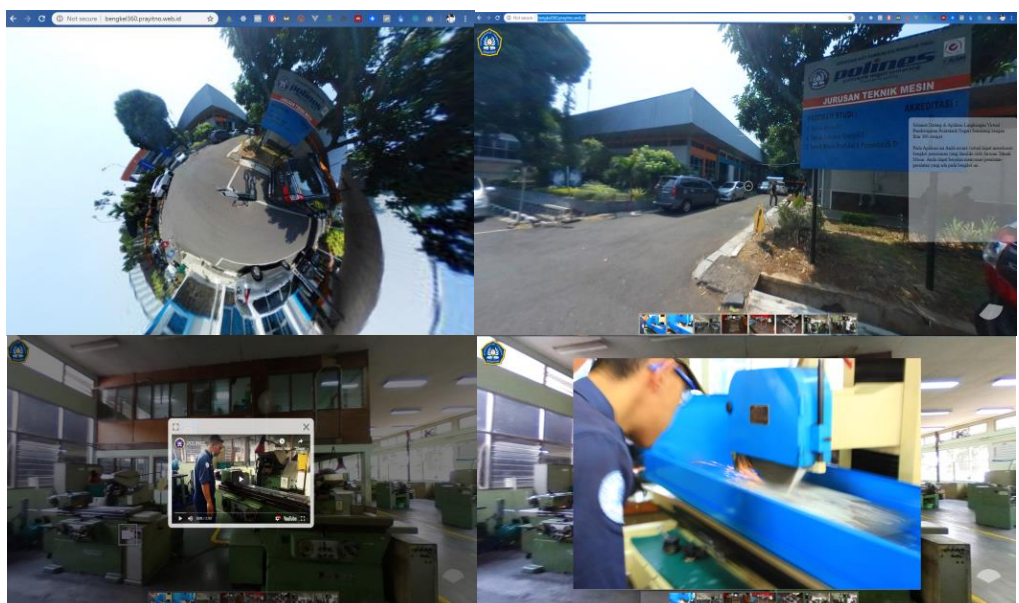

Figure 4. System Architecture: Upper left and upper right represent machinery workshop in virtual environment. Lower left shows machinery video tutorial in 360 environments. Lower right illustrates the machine details in pictures.

## V. Conclusion

This study has described the successful development of an educationally 360-degree virtual machinery workshop at vocational education material. Compared to handout learning material in classes, the 360-degree virtual learning environment in vocational education can be one of the innovative learning media that supports the teaching and learning process in the aspects of equipment introduction theory and workshop environment. This is based on augment reality and immersive technology that combines information in the real world into information that can be displayed through a website. Students and users can access information in the form of pictures and videos as if they were in an actual machinery workshop. In the future, we intend to study learning analytics for specific virtual learning environment to obtain more implied information about users learning behaviors.

## References

[1] J. Yip, S. H. Wong, K. L. Yick, K. Chan, and K. H. Wong, "Improving quality of teaching and learning in classes by using augmented reality video," *Comput. Educ.*, vol. 128, pp. 88–101, 2019.

[2] M. A. Rupp, K. L. Odette, J. Kozachuk, J. R. Michaelis, J. A. Smither, and D. S. McConnell, "Investigating learning outcomes and subjective experiences in 360-degree videos," *Comput. Educ.*, 2018.

[3] É. Ouellet, B. Boller, N. Corriveau-Lecavalier, S. Cloutier, and S. Belleville, "The Virtual Shop: A new immersive virtual reality environment and scenario for the assessment of everyday memory," *J. Neurosci. Methods*, vol. 303, pp. 126–135, 2018.

[4] R. Kleinert, N. Heiermann, R. Wahba, D. H. Chang, A. H. Hölscher, and D. L. Stippel, "Design, Realization, and First Validation of an Immersive Web-Based Virtual Patient Simulator for Training Clinical Decisions in Surgery," *J. Surg. Educ.*, vol. 72, no. 6, pp. 1131–1138, 2015.

[5] D. Nincarean, M. B. Alia, N. D. A. Halim, and M. H. A. Rahman, "Mobile Augmented Reality: The Potential for Education," *Procedia - Soc. Behav. Sci.*, vol. 103, pp. 657–664, 2013.

[6] C. M. Harrington *et al.*, "360° Operative Videos: A Randomised Cross-Over Study Evaluating Attentiveness and Information Retention," *J. Surg. Educ.*, vol. 75, no. 4, pp. 993–1000, 2018.

[7] I. Dubovi, S. T. Levy, and E. Dagan, "Now I know how! The learning process of medication administration among nursing students with non-immersive desktop virtual reality simulation," *Comput. Educ.*, vol. 113, pp. 16–27, 2017.

[8] T. Jeřábek, V. Rambousek, and R. Wildová, "Specifics of Visual Perception of the Augmented Reality in the Context of Education," *Procedia - Social and Behavioral Sciences*, vol. 159. pp. 598–604, 2014.

[9] W. Huang *et al.*, "Utility-oriented resource allocation for 360-degree video transmission over heterogeneous networks," *Digit. Signal Process. A Rev. J.*, vol. 84, pp. 1–14, 2019.

[10] A. Arora, L. Y. M. Lau, Z. Awad, A. Darzi, A. Singh, and N. Tolley, "Virtual reality simulation training in Otolaryngology," *Int. J. Surg.*, vol. 12, no. 2, pp. 87–94, 2014.

[11] T. Adão, L. Pádua, M. Fonseca, L. Agrellos, J. J. Sousa, and L. Magalhães, "A rapid prototyping tool to produce 360º video-based immersive experiences enhanced with virtual/multimedia elements," *Procedia Comput. Sci.*, vol. 138, pp. 441–453, 2018.

[12] E. Pastorelli and H. Herrmann, "A small-scale, low-budget semi-immersive virtual environment for scientific visualization and research," *Procedia Comput. Sci.*, vol. 25, pp. 14–22, 2013.

[13] D. Bhattacharjee, A. Paul, J. H. Kim, and P. Karthigaikumar, "An immersive learning model using evolutionary learning," *Comput. Electr. Eng.*, vol. 65, pp. 236–249, 2018.

[14] M. Novotný, J. Lacko, and M. Samuelčík, "Applications of multi-touch augmented reality system in education and presentation of virtual heritage," *Procedia Computer Science*, vol. 25. pp. 231–235, 2013.

[15] L. Rentzos, S. Papanastasiou, N. Papakostas, and G. Chryssolouris, *Augmented reality for human-based assembly: Using product and process semantics*, vol. 12, no. PART 1. IFAC, 2013.

[16] J.-G. Han, K.-W. Park, K.-J. Ban, and E.-K. Kim, "Cultural Heritage Sites Visualization System based on Outdoor Augmented Reality," *AASRI Procedia*, vol. 4, pp. 64–71, 2013.

[17] L. M. Dawborn-Gundlach *et al.*, "Enhancing pre-service teachers' concept of Earth Science through an immersive, conceptual museum learning program (Reconceptualising Rocks)," *Teach. Teach. Educ.*, vol. 67, pp. 214–226, 2017.

[18] Z. Lv, X. Li, and W. Li, "Virtual reality geographical interactive scene semantics research for immersive geography learning," *Neurocomputing*, vol. 254, pp. 1339–1351, 2017.